\documentstyle[12pt,procsla,epsf]{article}
\begin{document}
\def\lsim{{\buildrel < \over\sim}}
\def\gsim{{\buildrel > \over\sim}}
\def\to{\rightarrow}
\def\fb{~{\rm fb}}
\def\pb{~{\rm pb}}
\def\ev{\,{\rm eV}}
\def\kev{\,{\rm KeV}}
\def\mev{\,{\rm MeV}}
\def\gev{\,{\rm GeV}}
\def\tev{\,{\rm TeV}}
\def\wh{\widehat}
\def\wt{\widetilde}
\def\mhalf{m_{1/2}}
\def\gl{\wt g}
\def\q{$q$}
\def\qbar{$\bar{q}$}
\def\g{$g$}
\def\dc{$\delta_c$}
\def\als{\alpha_s}
\newcommand{\as}{{\ifmmode \alpha_S \else $\alpha_S$ \fi}}
\def\ttbs{\char'134}
\def\AmS{{\protect\the\textfont2
  A\kern-.1667em\lower.5ex\hbox{M}\kern-.125emS}}
\def\calN{{\cal N}}
\def\calO{{\cal O}}
\def\calP{{\cal P}}
\def\calL{{\cal L}}
\def\xbj{x_{\rm Bj}}
\def\xp{x_{\cal P}}
\def\cR{{\cal R}}
\catcode`@=11
\def\Biggg#1{\hbox{$\left#1\vbox to 22.5\p@{}\right.\n@space$}}
\catcode`@=12
\newcommand\refq[1]{$^{#1}$}
\newcommand\ind[1]{_{\rm #1}}
\newcommand\aopi{\frac{\as}{\pi}}
\newcommand\oot{\frac{1}{2}}
\newcommand\sinsthw{\sin^2(\theta_{\rm W})}
\newcommand\logmu{\log(\mu^2/\Lambda^2)}
\newcommand\Lfb{\Lambda^{(5)}}
\newcommand\Lfc{\Lambda^{(4)}}
\newcommand\Lf{\Lambda_5}
\newcommand\epem{\ifmmode e^+e^- \else $e^+e^-$ \fi}
\newcommand\mupmum{ \mu^+\mu^- }
\newcommand\bbar{b\bar{b}}
\newcommand\gamgam{\gamma\gamma }
\newcommand\ms{\ifmmode{\overline{\rm MS}}\else $\overline{\rm MS}$\ \fi}
\newcommand\Q[1]{_{\rm #1}}
\newcommand\pplus[1]{\left[\frac{1}{#1}\right]_+}
\newcommand\plog[1]{\left[\frac{\log(#1)}{#1}\right]_+}
\newcommand\sh{\hat{s}}
\newcommand\epbar{\overline\epsilon}
\newcommand\nf{\alwaysmath{{n_{\rm f}}}}
\newcommand\MSB{\ifmmode{\overline{\rm MS}}\else $\overline{\rm MS}$\ \fi}
\newcommand{\aem}{\alpha_{\rm em}}
\newcommand{\nlf}{\alwaysmath{{n_{\rm lf}}}}
\newcommand{\ep}{\epsilon}
\newcommand{\aop}{\frac{\as}{2 \pi}}
\newcommand{\Tf}{{T_{\rm f}}}
\newcommand{\mub}{\ifmmode \mu{\rm b} \else $\mu{\rm b}$ \fi}
\newcommand\alwaysmath[1]{\ifmmode #1 \else $#1$ \fi}
\newcommand{\TeV}{{\rm TeV}}
\newcommand{\GeV}{{\rm GeV}}
\newcommand{\MeV}{{\rm MeV}}
\newcommand{\LQCD}{\ifmmode \Lambda_{\rm QCD} \else $\Lambda_{\rm QCD}$ \fi}
\newcommand{\LMSB}{\ifmmode \Lambda_{\overline{\rm MS}} \else
          $\Lambda_{\overline{\rm MS}}$ \fi}
\newcommand{\qb}{\overline{q}}
\def\pp{\ifmmode p\bar{p} \else $p\bar{p}$ \fi}
\def\VEV#1{\left\langle #1\right\rangle}
\def\LMSb{\ifmmode \Lambda_{\rm \overline{MS}} \else
$\Lambda_{\rm \overline{MS}}$ \fi}
\def\ie{\hbox{\it i.e.}{}}      \def\etc{\hbox{\it etc.}{}}
\def\eg{\hbox{\it e.g.}{}}      \def\cf{\hbox{\it cf.}{}}
\def\etal{\hbox{\it et al.}}
\def\dash{\hbox{---}}
\def\abs#1{\left| #1\right|}   
\def\to{\rightarrow}
\def\d{{\rm d}}
\newcommand{\lra}{\leftrightarrow}
\newcommand{\la}{\langle}
\newcommand{\dd}{{\rm d}}
\newcommand{\PS}{{\rm PS}}
\newcommand{\pperp}{p_{\perp}}
\newcommand{\ra}{\rangle}
\def\vspaceinarray{\nonumber ~&~&~\\}

\newcommand{\Nc}{N_c}
\newcommand{\Nf}{N_f}
\def\LO{leading order }
\newcommand{\eps}{\epsilon}
\newcommand{\ve}{\varepsilon}
\newcommand\epb{\overline{\epsilon}}
\newcommand{\be}{\begin{equation}}
\newcommand{\ee}{\end{equation}}
\newcommand{\bea}{\begin{eqnarray}}
\newcommand{\eea}{\end{eqnarray}}
\newcommand{\beas}{\begin{eqnarray*}}
\newcommand{\eeas}{\end{eqnarray*}}
\def\abs#1{\left| #1\right|}
\def\Am{{\cal A}}
\def\nn{\nonumber}
\def\phys{{\rm phys}}
\def\ms{$\overline{{\rm MS}}$}
\def\limes#1{\mathrel{\mathop{\lim}\limits_{#1}}}
\def\arrowlimit#1{\mathrel{\mathop{\longrightarrow}\limits_{#1}}}
\def\mus#1#2{\left(-\frac{\mu^2}{s_{#1#2}}\right)^\varepsilon}
\def\qb{\bar{q}}
\def\Qb{\bar{Q}}

\relax
\def\ap#1#2#3{
        {\it Ann. Phys. (NY) }{\bf #1} (19#3) #2}
\def\app#1#2#3{
        {\it Acta Phys. Pol. }{\bf #1} (19#3) #2}
\def\ar#1#2#3{
        {\it Ann. Rev. Nucl. Part. Sci. }{\bf #1} (19#3) #2}
\def\cmp#1#2#3{
        {\it Commun. Math. Phys. }{\bf #1} (19#3) #2}
\def\cpc#1#2#3{
        {\it Comput. Phys. Commun. }{\bf #1} (19#3) #2}
\def\ijmp#1#2#3{
        {\it Int .J. Mod. Phys. }{\bf #1} (19#3) #2}
\def\ibid#1#2#3{
        {\it ibid }{\bf #1} (19#3) #2}
\def\jmp#1#2#3{
        {\it J. Math. Phys. }{\bf #1} (19#3) #2}
\def\jetp#1#2#3{
        {\it JETP Sov. Phys. }{\bf #1} (19#3) #2}
\def\ib#1#2#3{
        {\it ibid. }{\bf #1} (19#3) #2}
\def\mpl#1#2#3{
        {\it Mod. Phys. Lett. }{\bf #1} (19#3) #2}
\def\nat#1#2#3{
        {\it Nature (London) }{\bf #1} (19#3) #2}
\def\np#1#2#3{
        {\it Nucl. Phys. }{\bf #1} (19#3) #2}
\def\npsup#1#2#3{
        {\it Nucl. Phys. Proc. Sup. }{\bf #1} (19#3) #2}
\def\pl#1#2#3{
        {\it Phys. Lett. }{\bf #1} (19#3) #2}
\def\pr#1#2#3{
        {\it Phys. Rev. }{\bf #1} (19#3) #2}
\def\prep#1#2#3{
        {\it Phys. Rep. }{\bf #1} (19#3) #2}
\def\prl#1#2#3{
        {\it Phys. Rev. Lett. }{\bf #1} (19#3) #2}
\def\physica#1#2#3{
        { Physica }{\bf #1} (19#3) #2}
\def\rmp#1#2#3{
        {\it Rev. Mod. Phys. }{\bf #1} (19#3) #2}
\def\sj#1#2#3{
        {\it Sov. J. Nucl. Phys. }{\bf #1} (19#3) #2}
\def\zp#1#2#3{
        {\it Zeit. Phys. }{\bf #1} (19#3) #2}
\def\tmf#1#2#3{
        {\it Theor. Math. Phys. }{\bf #1} (19#3) #2}

\vspace*{0.6cm}

\begin{flushright}
{\large ETH--TH/97--30}\\
{\rm September 1997\hspace*{.5 truecm}}\\
\end{flushright}

\vspace*{3truecm}

\begin{center}
{\Large \bf 
HARD DIFFRACTIVE SCATTERING 
\footnote{Talk given at the Ringberg Workshop
 New Trends in HERA Physics, 25 - 30 May 1997;\,
E-mails:
 kunszt@itp.phys.ethz.ch} }\\[0.5cm]
 {\large 
Zoltan ~Kunszt}\\[0.15 cm]
{\it  Institute of Theoretical Physics, ETH, Z\"urich, Switzerland.}\\[0.15cm]
\end{center}
\vspace*{2truecm}
\begin{abstract}
\noindent\small 
I discuss  hard diffractive scattering  in the
framework of perturbative QCD and Regge-parametrization.

\end{abstract}
\vspace*{2cm}


\section{Diffractive deep inelastic scattering}
\subsection{Kinematics}
Measurements of deep inelastic diffractive scattering at HERA~\cite{ZEUS,H1,H1new}
provided us with some of the most interesting experimental results
which allow one to test new theoretical ideas on
diffractive processes in an unexplored regime. The  data 
triggered much theoretical work. The concept of diffractive
parton number densities
 (or differential fracture functions)\cite{TRENVEN,SOPER}
 with standard DGLAP evolution
appears to be the key ingredient of the QCD description.
It was  conjectured that these new types of parton number densities
factorize similarly to the parton number densities
of non-diffractive deep inelastic scattering and satisfy the same
Altarelli-Parisi evolution equation.
The Ingelman-Schlein model\cite{IS} emerges subsequently by assuming
 Regge-behaviour\cite{DL} for the $\xp$ and $t$ dependence.

Diffractive deep inelastic scattering is defined by restricting our studies into
final states when the hadrons can be classified into a
high mass ($M_X$) and a low mass ($M_Y$) group with large rapidity gap  between them.
Denoting the four-momenta of virtual photon, proton and electron  with $q,p$ and
$k$ and the four-momenta of the high mass and low mass 
hadronic system with $p_X$ and $p_Y$, respectively, the measured differential
cross-section can be written in terms of the
 diffractive structure  function $F_2^D$ as
\begin{equation}
\frac{d\sigma^D}{dx dQ^2 dx_{\calP}d\abs{t}}
=\frac{4\pi\alpha^2}{x Q^4}\left(1-y+\frac{y^2}{2}\right)
\frac{d F_2^D}{d\xp d{\abs{t}}}(x,Q^2,x_{\calP})
\end{equation}
where  
\begin{eqnarray}
Q^2&=&-q^2\,,\quad y=\frac{q.p}{k.p}\,,\quad x=\frac{Q^2}{2p.q}\,,\quad
x_{\calP}=\frac{q.p-q.p_y}{q.p}\approx \frac{Q^2+M^2_X}{Q^2+W^2}\,, 
\nonumber\\
\beta&=&\frac{x}{x_{\calP}}=\frac{Q^2}{2q.p-2q.p_y}\approx \frac{Q^2}{Q^2+M^2_X}\,,\quad 
W^2=(q+p)^2\,,\quad t=(p-p_Y)^2
\end{eqnarray}
Requiring small $M_Y$ and $x_{\calP}$ the rapidity gap is inevitable. 
For example the H1 collaboration required $M_Y<1.6 \gev^2$ and $x_{\calP}<0.05$.
In the H1 measurement, the momentum transfer from the proton to the $Y$-system $t=(p-p_Y)^2$ is not 
measured  but is restricted to the small interval $\abs{t_{\rm
min}}<\abs{t}<1.0 \gev^2$.
  The ZEUS collaboration collected also data when the
$Y$ system was identified with a fast forward proton with a measured value 
of $t$. The longitudinal cross-section was assumed to be zero.

\subsection{Diffractive parton number densities}
The   factorization theorem of deep inelastic
scattering allows the decomposition
\begin{equation}
F_2(x,Q^2)=\sum_a\int dx^{\prime} f_{a/p}(x^{\prime},\mu)
\hat{F}_{2a}(x/x^{\prime},Q^2,\mu^2)
\end{equation}
It is natural to ask whether similar decomposition 
remain valid also for diffractive deep inelastic scattering giving
\begin{equation}
\frac{d F_2}{dx_{\calP}d \abs{t}}(x,Q^2,x_{\calP},t)=\sum_a\int dx^{\prime}
\frac{df_{a/p}}{dx_{\calP}d \abs{t}}(x^{\prime},\mu,x_{\calP},t)
\hat{F}_{2a}(x/x^{\prime},Q^2,\mu^2)
\end{equation}
where  in both equations $\hat{F}_2$ denotes
 the finite hard scattering contribution. 
The dependence on $t$ and $x_{\calP}$  is completely absorbed  in the
diffractive parton number densities.
Berera and Soper\cite{SOPER} pointed out that the  operator definition of the parton
number densities can be generalized for diffractive processes.
The  quark parton number density for example defined as
\begin{eqnarray}
 &&f_{q/p}(x,\mu)=
\frac{1}{4\pi}\sum_{s,X} \int dy^{-} e^{ix p^+y^-}\nonumber \\
&&\qquad
<p,s|\tilde{\bar{q}}(0,y^-,\vec{0})|X>
\gamma^+<X|\tilde{q}(0,0,0)|p,s>
\end{eqnarray}
is modified in the case of diffractive densities as
\begin{eqnarray}
 &&\frac{df_{q/p}}{dx_{\calP}d \abs{t}}(\beta,\mu,x_{\calP},t)=
\frac{1}{4\pi}\sum_{s,s^{'},X} \int dy^{-} e^{i\beta p^+y^-}\nonumber \\
&&\qquad <p,s|\tilde{\bar{q}}(0,y^-,\vec{0})|
p^{'},s^{'};X>\gamma^+<p^{'},s^{'};X|\tilde{q}(0,0,0)|p,s>
\end{eqnarray}
where $X$ indicates all possible final states with a  proton  of
momentum $p'$ and polarization $s'$
 and
$\tilde{q}$ is  the color singlet  quark operator given in terms of quark and
gluon fields  as 
\begin{equation} 
\tilde{q}(0,y^-,\vec{0})={\cal P}e^{ig\int_{y^-}^{\infty}dx^- A^+_c(0,x^-,\vec{0})
T_c}q(0,0,0) 
\end{equation}
where $T_c$ is the color matrix and ${\cal P}$ demands path ordering in color space.
In fig.1 the factorization in DIS is shown graphically. 
The crucial observation is that
the ultraviolet structure of the bilocal operators
defining the non-diffractive   and  diffractive parton
number densities are the same
therefore they fulfill the same  renormalization group equation.
As a result,  the diffractive parton number densities at fixed values of $x_{\calP}$ and
$t$ (assuming $t$ is small)  also have to fulfill the DGLAP evolution equation.
The apparent difference in the $Q^2$ behaviour
of the DIS and diffractive DIS data must have its origin in
the dramatically different initial parton distributions.
\begin{figure}[htb]
\epsfysize=6truecm
\centerline{\epsffile{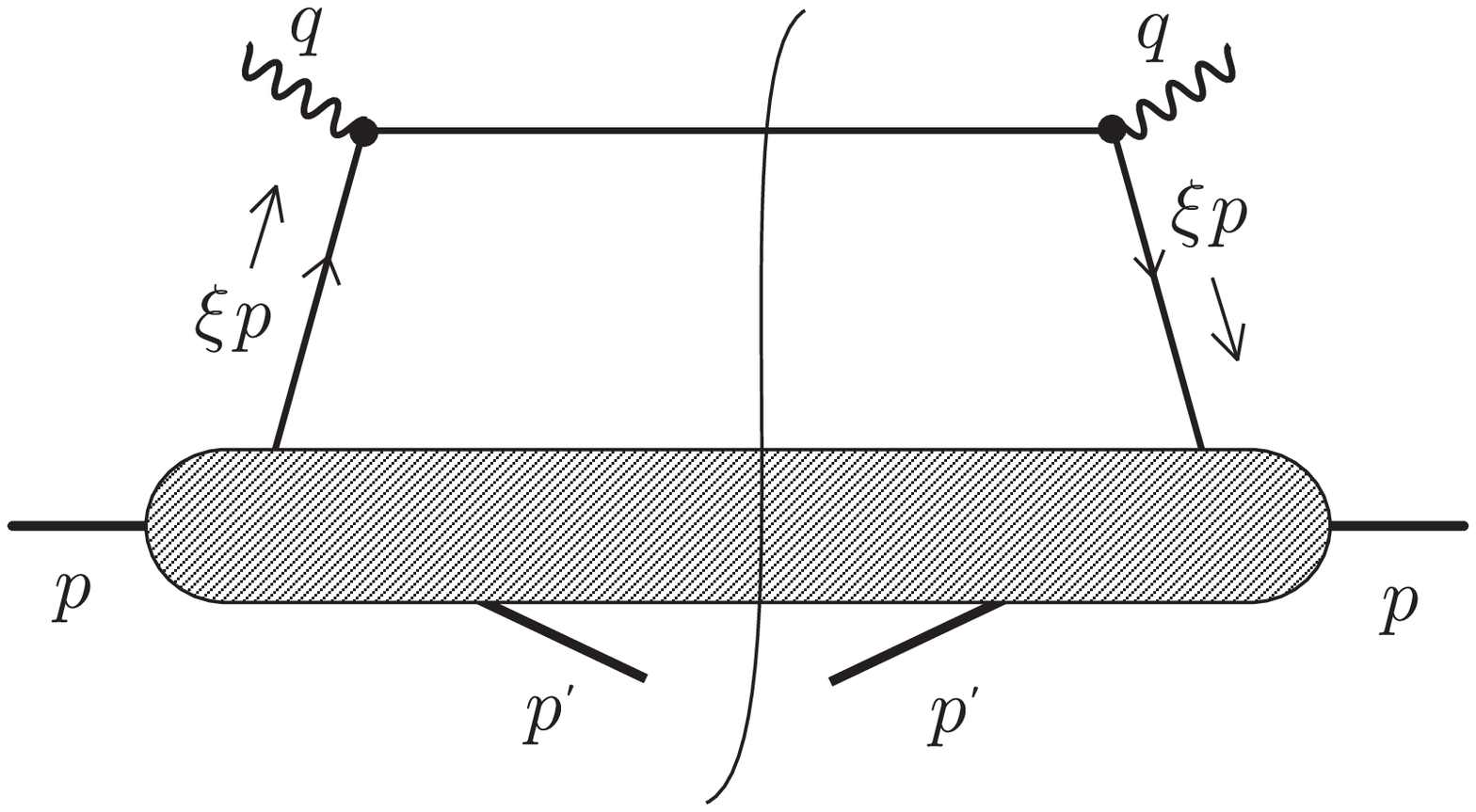}}
\vskip6pt
\baselineskip=13pt
\fcaption{ Diffractive factorization in DIS. }
\vskip 0.1cm
\end{figure}
Factorization theorem with  evolution
equation
has been suggested by Trentadue and Veneziano\cite{TRENVEN}
in the context of deep inelastic scattering
with one observed final state hadron allowing also target fragmentation region.
 In the QCD improved parton model  collinear
singularities given by
 low $p_T$  gluons 
emitted into the target fragmentation region  can not be absorbed
into the standard fragmentation and structure functions and a new
non--perturbative  function, the so called fracture function, had  to be introduced.
In  diffractive deep inelastic scattering
the observed forward proton is in the 
target fragmentation region. Therefore 
  the fracture function formalism applies.
Berera and Soper pointed out\cite{SOPER} that
if the forward proton  transverse momentum remains unintegrated and
is required to be small the diffractive parton number densities can be
identified with the fracture functions.
 With explicit calculation Graudenz\cite{GRAUDENZ} has shown that the
factorization theorem for fracture functions is fulfilled in next-to-leading order
accuracy. It is generally believed that it remains valid   
for diffractive  deep inelastic scattering  to all orders in
perturbation theory.

\subsection{Regge factorization and parametrization}

As we have seen above both the standard parton number densities
and the diffractive parton number densities are defined 
as matrix element of bilocal operator between proton states and
they are given by 'soft physics', they 
can not be  calculated in perturbation theory.
The parametrization of the initial distribution at some $Q^2_0$ scale,
however, can be motivated by phenomenological models.
 For diffractive scattering  Regge theory is the most successful framework.
Assuming the dominance of the Pomeron trajectory one obtains
\be
\frac{df_{q/p}}{dx_{\calP}d \abs{t}}(\beta,\mu,x_{\calP},t)=
\frac{\abs{\gamma(t)}^2}{8\pi^2}\xp^{-2\alpha_{\calP}(t)} f_{a/{\calP}}(\beta,t;\mu)
\ee
where $\alpha_{\calP}(t)$ is the trajectory function and $\gamma (t)$ is its coupling to
the proton. From fits to  hadronic cross-sections\cite{DL} one obtains $\alpha_{\calP}\approx 1.08$.
The function $f_{a/{\calP}}(\beta,t;\mu)$ is called the parton distribution of the Pomeron
since formally the cross-section of diffractive deep inelastic scattering is given
by folding the hard scattering cross-section with this function. This description
was suggested by Ingelman and Schlein using the notion of 'parton constituents of the Pomeron'.
The latter concept should be treated with care since the Pomeron can not be interpreted as a
particle emitted by the proton before the parton distribution was probed.
One of the most significant qualitative  consequences of the assumption of the Pomeron parametrization
is the factorization of the $\xp$ and $t$ dependence. 
The  HERA  diffractive structure
function data, however, show  a modest amount of factorization breaking
\cite{H1,ZEUS,H1new}. It 
could, however, 
 be accommodated by  invoking a sum over Regge trajectories, each
with a different intercept and structure function:
\be
 F_2^{D}(\beta,Q^2; \xp,t) = \sum_{\cR}\;
 F_{\cR}(t) \xp^{-2\alpha_{\cR}(t)+1}\;F_2^{\cR}(\beta,Q^2)\; ,
\label{eq:reggeons}
\ee
which would yield an effective power of the $\xp$ dependence which depends on $\beta$
but is approximately independent of $Q^2$. Integrating over  
$t$  and small bins of $\xp$ what is measured is
a linear combination of $F_2^\cR$ structure functions or, equivalently,
the parton distributions in an effective colour singlet target:
\be
\int d \xp dt \; F_2^{D}(\beta,Q^2; \xp,t) = \sum_{\cR}\;
A_\cR \;F_2^{\cR}(\beta,Q^2) = \beta \sum_q e_q^2\; \sum_{\cR}\;
 q_{\cR}(\beta,Q^2)\; ,
\ee
where the coefficients $A_\cR$ are independent of $\beta$ and $Q^2$.
Since the DGLAP equations are {\it linear} in the parton distributions,
the $Q^2$ evolution of the integrated  structure function $F_2^D$ 
should also be calculable perturbatively.
The $\xp$ dependence 
was fitted by treating  $\alpha_{\calP}(0)$ as a free parameter
and the H1 and ZEUS collaboration have found that its value is somewhat
larger than the soft Pomeron values.
The value of a recent fit obtained by H1 is
$\alpha_{{\cal P}}\approx 1.20 \pm 0.04$. The values obtained
by H1 and ZEUS are consistently between the values of the
intercept of the soft and hard (BFKL) Pomeron. 
\begin{figure}[ht!]
\vskip -2truecm
\epsfysize=8truecm
\centerline{\epsffile{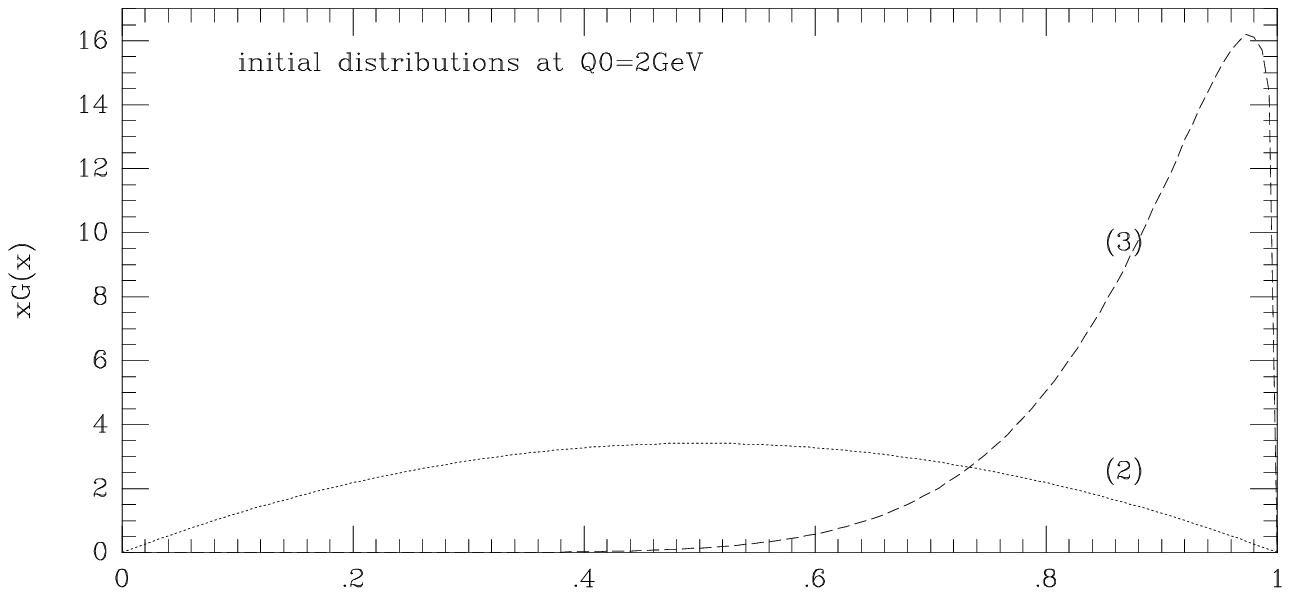}}
\vskip 6.8cm
\baselineskip=13pt
\fcaption{Starting parton distributions and their evolved values }
\end{figure}

Various fits for the parton distributions in the
`Pomeron'   $f_{q/\calP}(x,\mu^2)$  have been proposed,
ranging from the two extremes of mainly gluons to mainly quarks 
\cite{H1,GS,ROMKS,Alvero}).
 Already  the early data
favoured the hard gluon parametrization. From the  recent H1 analysis of the
1994 data\cite{H1new} one can conclude that
 the parametrization which only assumes quarks at $Q^2=3\gev^2$ is 
excluded.
Since the diffractive parton number densities do not have momentum sum rule
the  overall normalization of the quark singlet $\Sigma(x,\mu^2)$ and the gluon
$g(x,\mu^2)$ densities have to also be   fitted from the
data. The normalization, however, will not change
with $Q^2$ evolution. 
It is the {\it product} of $f_{q/\calP}$ and $f_\calP$
that appears in the expression for the structure function, therefore it is  convenient to
simply impose a momentum sum rule on the parton distributions and
absorb an overall normalization $\calN$ into $f_\calP$. 
In fig.~2 the initial parton distributions are shown at $\mu^2=4\gev^2$
for hard gluon (3) no-gluon (1) and soft gluon (2) fits.

There has been some warning\cite{bartels}
 that one can not accept the fitted gluon distribution if it is
peaked at $x=1$ since the higher twist contributions become important
in the range $(1-x)\mu^2 < 1\gev^2$. This criticism, however, should concern
only the measured data points.
If  the data data points in this range are excluded from the fit
the obtained densities can be meaningfully calculated using the
DGLAP evolution equation also in this range. 
One can easily see that the densities in this range can not be directly
related to measurable quantities without adding also higher twist terms.

Finally I note that a number of  models have been investigated in the literature
to calculate the diffractive deep inelastic structure function at some fixed scale.
Interesting  semi-classical
 description  has  been developed by Buchmueller and Hebbeker\cite{buchmueller,mcdermott}
consistent with the aligned jet parton picture\cite{mcdermott,bjaligned}. The Pomeron is soft and in leading order
the intercepts is $\alpha_{\calP}(0)=1.0$. Models with simple two gluon exchanges 
have also been considered~\cite{bartels}.

\section{Hard diffractive processes in hadron-hadron scattering}
Assuming the validity of factorization theorem also
for diffractive hard scattering processes
using the DIS value of the densities
  the cross-section values in hadron-hadron
collisions can be calculated  by using the standard formula
\be
\frac{d\Delta\sigma^{\rm SD}}{d\xp dt}=
\sum_{a,b}\int f_{a/p}(x_a,\mu)\frac{df_{a/p}}{d\xp dt}
(x_b,\mu,\xp,t)\Delta\hat{\sigma}(x_ax_bs,..)
\ee
Here the label SD makes reference to single diffractive production. The generalization of this
formula for double diffractive scattering is trivial. Data on single diffractive
jet production has been obtained first by UA8\cite{UA8}. Its comparison
with the theory was not clear because of  high values of $t$.
 The validity of the factorization theorem, however, is 
debated\cite{frankfurt}. In particular it is expected that it may be strongly violated.
The soft exchanges before the hard scattering do not
cancel if hadrons are observed in the target fragmentation region
because  not all  the final  states containing the hard-scattered object (jet
or $W$-boson or Higgs boson etc.)  are  summed over. 
Interaction between spectator partons can lead to extra particles to fill the
rapidity gap. It has been suggested to correct the prediction of the naive
application of the factorization theorem with
the so called survival probability of the rapidity gap\cite{bjorkensur,soper97}. 
The parton model is predicted to overestimate the cross-sections of hard
diffractive scattering by an amount given by such a survival probability.
To gain experimental information on this relevant question one may proceed 
by  comparing the naive predictions with  the
data. 
 The cleanest process to study
would appear to be weak boson production at the Tevatron $p \bar p$
collider. The naive prediction  was first
calculated  before the HERA data  in Ref.~\cite{BRUNI} neglecting possible
effects of $Q^2$ evolution.  It has been found
that the single diffractive component of the total $W$ cross section
could be as large as $20\%$.
It is convenient to study integrated  `single diffractive' events
by $x_\calP < 0.1$, integrating over all $t$ and to normalize the
result to the total $W$-production cross-section. In practice, of course,
the events are defined by rapidity gaps of a certain minimum size,
and therefore the observed diffractive cross section must be corrected
to the theoretical prediction based on, say, $x_\calP < 0.1$ using
a Monte Carlo simulation~\cite{GOULIANOS}.
Using the various fits discussed above, it is straightforward to
reevaluate the single diffractive $W$ cross section with $Q^2$ evolution and 
with updated values for the Pomeron flux factors. Two recent analysis\cite{ROMKS,Alvero}
have found values in the range
\be
\frac{\Delta\sigma_W^{\rm SD, th}}{\sigma_W} \approx 
 4\%\dash 6\%.
\ee
The important point to note here is that the  predictions coming from different fits
are quite similar. The single diffractive $W$ cross
section at the Tevatron samples the quarks  at
$\langle x_{q/\calP} \rangle \sim 0.4$. At low $Q^2$ the quark
distributions at this $x$
are constrained by the HERA $F_2^D$ data to be roughly the same. As
$Q^2$ increases the  distributions diverge, reflecting the
quantitatively different
gluon contributions to the DGLAP evolution. However at $Q^2 \sim
10^{4}$~GeV$^2$, the relevant value for $W$ production, the difference
between the quarks in the three models is still not very large, 
and the predictions for $\sigma^{SD}(W)$ are correspondingly similar since they
are directly related to the HERA data.
In recent measurements the  CDF Collaboration has found a smaller ratio
\be
\frac{\Delta\sigma_W^{\rm SD, th}}{\sigma_W} \approx 
1.55\pm .55
\ee
The surviving probability of the factorized result is about $S\approx 20-30\%$ .
Similar analysis for inclusive jet production gave even
smaller  value $S \approx 10\%$.
These results indicate that
 the use of the 
(invalid) factorization theorem very likely strongly overestimates
 the cross-section values  of diffractively produced heavy quarks and 
Higgs bosons at LHC.
It would be important to develop some theoretical model in which
the survival probability could be estimated.


\section{Acknowledgments}
I thank Dave Soper and James Stirling for helpful discussions. I also thank
F. Hautmann for reading the manuscript and
  Bernd Kniehl for organizing an interesting  workshop.

\end{document}